\documentclass[preprint,showpacs,preprintnumbers,amsmath,amssymb]{revtex4}

\usepackage{graphicx}
\usepackage{dcolumn}
\usepackage{bm}

\newcommand{\qed}{\nobreak \ifvmode \relax \else
      \ifdim\lastskip<1.5em \hskip-\lastskip
      \hskip1.5em plus0em minus0.5em \fi \nobreak
      \vrule height0.75em width0.5em depth0.25em\fi}

\begin{document}

\preprint{}

\title{Bures and Hilbert-Schmidt $2 \times 2$ Determinantal Moments}
\author{Paul B. Slater}%
\email{slater@kitp.ucsb.edu}
\affiliation{%
University of California, Santa Barbara, CA 93106-4030\\
}%
\date{\today}

\begin{abstract}
We seek to gain insight into the nature of the 
determinantal moments $\left\langle \left\vert \rho^{PT}\right\vert^{n}\left\vert \rho\right\vert^{k}\right\rangle^{Bures}$ of generic (9-dimensional) two-rebit and (15-dimensional) two-qubit systems ($\rho$), 
$PT$ denoting partial transpose. Such information--as it has proved to be in the Hilbert-Schmidt counterpart--should  be useful, employing 
probability-distribution reconstruction (inverse) procedures, in obtaining Bures $2 \times 2$ separability probabilities. The (regularizing) strategy we first adopt is to plot the ratio of numerically-generated (Ginibre ensemble) estimates of the Bures moments to the corresponding (apparently) exactly-known  Hilbert-Schmidt moments ({\it J. Phys. A}, {\bf{45}}, 095305 [2012]). Then, through a combination of symbolic and numerical computations, we obtain strong evidence as to the exact values (and underlying patterns) of certain Bures determinantal moments. In particular, the first moment (average)
of $|\rho^{PT}|$ (where $|\rho^{PT}| \in [-\frac{1}{16},\frac{1}{256}]$) for the two-qubit systems is, remarkably,  $-\frac{1}{256} = -2^{-8} \approx -0.00390625$. The analogous value for the two-rebit systems is 
$-\frac{2663}{860160} = -\frac{2663}{2^{13} \times 3 \times 5 \times 7} \approx -0.00309594$. While $\frac{\left\langle \left\vert \rho^{PT}\right\vert ^{n}\left\vert \rho\right\vert
^{k}\right\rangle}{\left\langle \left\vert \rho\right\vert ^{k}\right\rangle}$ in the Hilbert-Schmidt case is the ratio of $3 n$-degree polynomials in $k$, it appears to be the ratio of $5 n$-degree polynomials in $k$ in the Bures case.
\end{abstract}


\maketitle
\section{Introduction}
{\.Z}yczkowski, Horodecki, Sanpera and Lewenstein raised and investigated--in two-qubit and qubit-qutrit settings--the clearly non-trivial, fundamental quantum-information-theoretic question of what proportion of generic bipartite quantum states  is separable \cite{ZHSL}. Over the past several years, we had--employing Bures \cite{szBures,ingemarkarol} and Hilbert-Schmidt \cite{szHS,ingemarkarol} measures on the space of such states--examined this question using several distinct approaches--but with no definitive resolutions 
\cite{slaterA,slaterC,slaterJGP,slaterPRA,pbsCanosa,slater833,JMP2008,ratios}.
The problem appears to be particularly challenging, in large part, due to the high-dimensionality of the relevant spaces--15- and 35-dimensions in the generic two-qubit and qubit-qutrit cases, respectively. 

Recently, we have reported substantial advances---in the two-qubit ($2 \times 2$) case, employing  Hilbert-Schmidt (HS) 
measure--using a (somewhat indirect, but nevertheless effective) multi-stage approach based on, first, obtaining formulas for the HS moments of $4 \times 4$ density matrices ($\rho$) and of their partial transposes \cite{MomentBased,Hypergeometric,ConciseFormula}. (For example, the evidence derived--though not yet fully rigorous--is compelling that the Hilbert-Schmidt separability probabilities of generic [9-dimensional] two-rebit states is $\frac{29}{64}$, of [15-dimensional] two-qubit states is $\frac{8}{33}$, and of [27-dimensional] 
two-quater(nionic)bit states is $\frac{26}{323}$.)
It is the objective of this study to investigate and evaluate the possibility of fully adopting this apparently successful approach to the Bures case, as well. We are able to report some interesting progress in this direction, but still fall considerably short of such a long-range objective.
We will now briefly discuss these recent (Hilbert-Schmidt-based) advances, that motivate our present quest to, additionally, find Bures analogues. 

In \cite[App. D.6]{MomentBased}, formulas (not yet rigorously established) were developed  by Dunkl 
for the HS {\it bivariate} product determinantal moments 
$\left\langle \left\vert \rho^{PT}\right\vert^{n}\left\vert 
\rho\right\vert^{k}\right\rangle^{HS}_{\alpha} ,k,n=0,1,2,3,\ldots,$--$PT$ denoting the partial transpose. Here, $\alpha$ is a Dyson-index-like parameter, with, in particular, $\alpha=1$ denoting the two-qubit systems and $\alpha=\frac{1}{2}$, the two-rebit systems. These formulas then served as input   (using the induced {\it univariate} moments 
$\left\langle \left\vert \rho^{PT}\right\vert^{n} \right\rangle^{HS}_{\alpha}$  and 
$\left\langle \left\vert \rho^{PT}\right\vert^{n}\left\vert 
\rho\right\vert^{n}\right\rangle^{HS}_{\alpha}$ \cite{MomentBased,Hypergeometric}) to a (Legendre-polynomial-based) probability-distribution reconstruction procedure \cite{Provost}. This provided estimates 
(and eventually--{\it via} a hypergeometric-based expression generated--exact results) for Hilbert-Schmidt $2 \times 2$ separability probabilities 
as a function of $\alpha$.

In sec.~\ref{Plots} we will compare--graphically--Bures and Hilbert-Schmidt 
$2 \times 2$ determinantal moments for generic rebit and qubit systems,
while in sec.~\ref{Exact} we will begin--what appears to be a laborious process--of obtaining exact results pertaining to  Bures determinantal moments.
\section{Bures/Hilbert-Schmidt Determinantal Moment Ratio Plots} \label{Plots}
In the present study, we first employed the Ginibre ensemble procedures detailed in \cite{osipov} to generate 59,800,000  (9-dimensional) generic two-rebit  and (15-dimensional) generic two-qubit density matrices distributed according to Bures measure. Then, using high-precision arithmetic, we obtained estimates of the first 500 moments of the determinant $|\rho^{PT}|$--the nonnegativity of this determinant being necessary and sufficient for separability in the $2 \times 2$ case 
\cite{augusiak}. 

\begin{figure}
\includegraphics{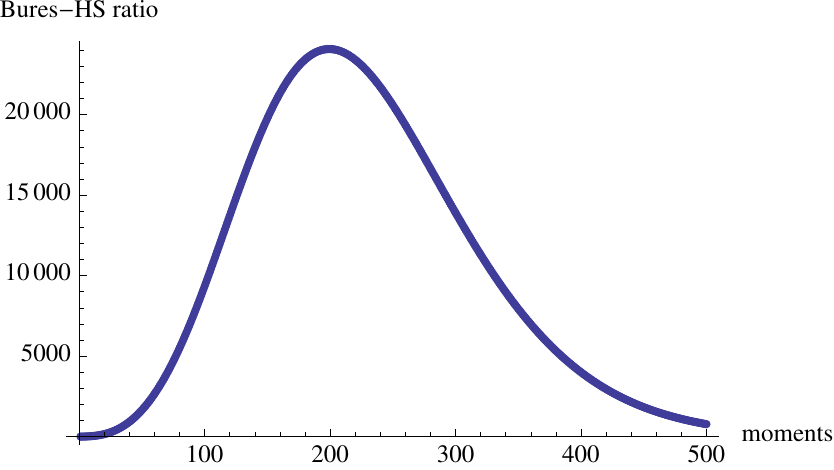}
\caption{\label{fig:RatioTwoRebitMoments}Ratio of numerically-generated Bures moments $\left\langle \left\vert \rho^{PT}\right\vert^{n} \right\rangle^{Bures}_{2-rebits}$--based upon 59,800,000 random realizations--to corresponding exact Hilbert-Schmidt moments}
\end{figure}

\begin{figure}
\includegraphics{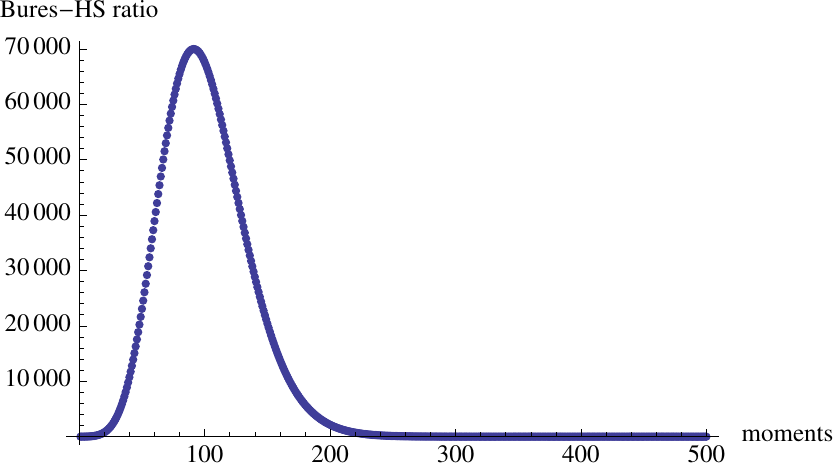}
\caption{\label{fig:RatioTwoQubitMoments}Ratio of numerically-generated Bures moments $\left\langle \left\vert \rho^{PT}\right\vert^{n} \right\rangle^{Bures}_{2-qubits}$--based upon 59,800,000 random realizations--to corresponding exact Hilbert-Schmidt moments}
\end{figure}

In Figs.~\ref{fig:RatioTwoRebitMoments} and \ref{fig:RatioTwoQubitMoments} we plot the ratios of these estimates to the exactly-known Hilbert-Schmidt moments using one of the formulas presented
 in \cite[App. D.6]{MomentBased}, that is
\begin{gather*} \label{nequalzero}
\left\langle \left\vert \rho^{PT}\right\vert ^{n}\right\rangle^{HS}_{\alpha} =\frac
{n!\left(  \alpha+1\right)  _{n}\left(  2\alpha+1\right)  _{n}}{2^{6n}\left(
3\alpha+\frac{3}{2}\right)  _{n}\left(  6\alpha+\frac{5}{2}\right)  _{2n}}\\
+\frac{\left(  -2n-1-5\alpha\right)  _{n}\left(  \alpha\right)  _{n}\left(
\alpha+\frac{1}{2}\right)  _{n}}{2^{4n}\left(  3\alpha+\frac{3}{2}\right)
_{n}\left(  6\alpha+\frac{5}{2}\right)  _{2n}}~_{5}F_{4}\left(
\genfrac{}{}{0pt}{}{-\frac{n-2}{2},-\frac{n-1}{2},-n,\alpha+1,2\alpha
+1}{1-n,n+2+5\alpha,1-n-\alpha,\frac{1}{2}-n-\alpha}%
;1\right) .
\end{gather*}
(Because of the occurrence of the term $1-n$ in the denominator, it is necessary to replace the
$_{5}F_{4}$-sum by $1$ to obtain the correct value when $n=1$.) 
This graphic approach might be regarded as a form of regularization of our numerical procedures, with the resulting ratios plotted in 
Figs.~\ref{fig:RatioTwoRebitMoments} and \ref{fig:RatioTwoQubitMoments} being of far less extreme magnitudes than the numerical estimates of the Bures moments themselves.


Since the lower-order Bures moments dominate the Hilbert-Schmidt ones in absolute value, it would appear that it would-attractively--take fewer Bures moments (if known exactly) than those several thousand Hilbert-Schmidt ones employed in \cite{MomentBased,Hypergeometric} to obtain comparable accuracies in the probability-distribution reconstruction procedure \cite{Provost}.

\begin{figure}
\includegraphics{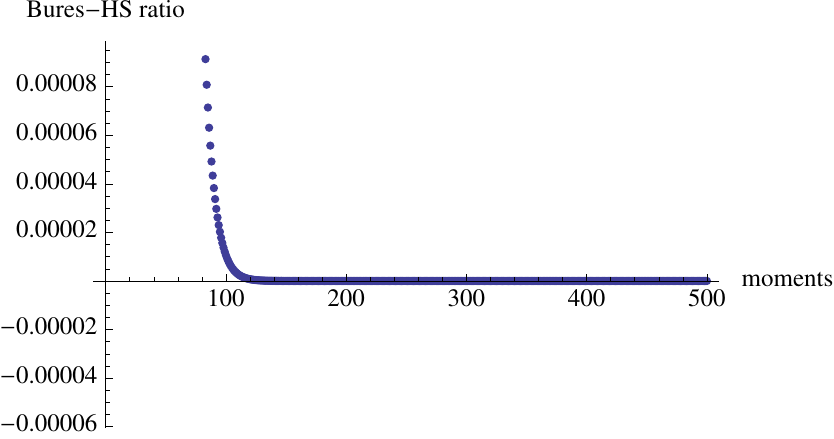}
\caption{\label{fig:RatioTwoRebitMomentsBalanced}Ratio of numerically-generated Bures moments $\left\langle (  |\rho^{PT}| |\rho|)^{n}  \right\rangle^{Bures}_{2-rebits}$--based upon 17,200,000 random realizations--to corresponding exact Hilbert-Schmidt moments. For $n=1$, the HS moment is zero.}
\end{figure}

\begin{figure}
\includegraphics{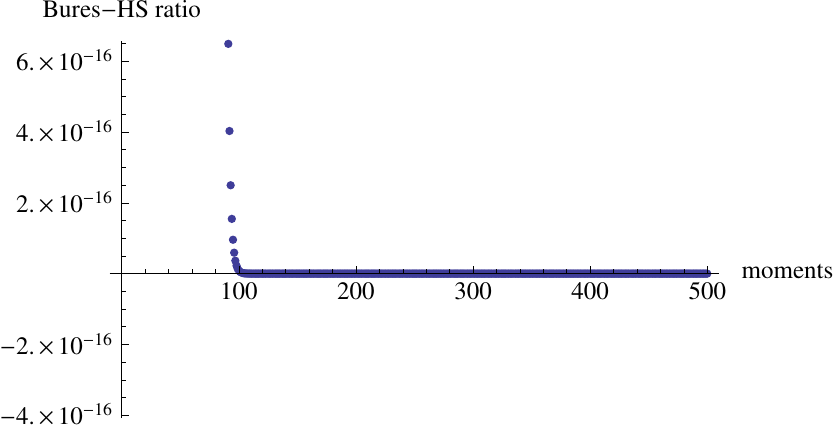}
\caption{\label{fig:RatioTwoQubitMomentsBalanced}Ratio of numerically-generated Bures moments $\left\langle (  |\rho^{PT}| |\rho|)^{n} \right\rangle^{Bures}_{2-qubits}$--based upon 17,200,000 random realizations--to corresponding exact Hilbert-Schmidt moments}
\end{figure}

Further, again using Ginibre ensembles \cite{osipov}, 
we obtained estimates of the first 
500 moments of the "balanced" variable $|\rho^{PT}| |\rho|$. 
(The range of this variable is $[-\frac{1}{110592}, \frac{1}{256^2}]$ 
\cite[sec. 3.1]{MomentBased}.) In 
Figs.~\ref{fig:RatioTwoRebitMomentsBalanced} and 
\ref{fig:RatioTwoQubitMomentsBalanced} we  plot the ratios of these estimates to the exactly-known corresponding Hilbert-Schmidt moments \cite[App. D.6]{MomentBased}, given by 
\begin{gather*} \label{nequalk}
\left\langle (|\rho|  |\rho^{PT}|)^{n} \right\rangle^{HS}_{\alpha} \\
=\frac{\left(  2n\right)  !\left(  1+\alpha\right)  _{2n}\left(
1+2\alpha\right)  _{2n}}{2^{12n}\left(  3\alpha+\frac{3}{2}\right)
_{2n}\left(  6\alpha+\frac{5}{2}\right)  _{4n}}~_{4}F_{3}\left(
\genfrac{}{}{0pt}{}{\ -n,\alpha,\alpha+\frac{1}{2},-4n-1-5\alpha
}{-2n-\alpha,-2n-2\alpha,\frac{1}{2}-n}%
;1\right)  .
\end{gather*}
(It is interesting to note, in this Hilbert-Schmidt setting, that the first moment of the balanced variable for the two-rebit systems [$\alpha= \frac{1}{2}$], given by the formula is zero. For the two-qubit systems [$\alpha =1$], however, the analogous 
first moment is nonzero, that is, $-\frac{1}{4576264}$.)


\begin{figure}
\includegraphics{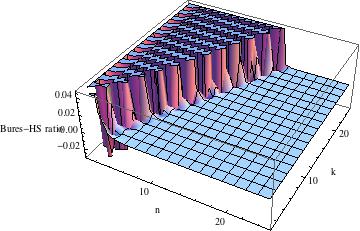}
\caption{\label{fig:RatioTwoRebitMoments2D}Ratio of numerically-generated {\it bivariate} Bures moments $\left\langle 
\left\vert \rho^{PT}\right\vert^{n} \left\vert \rho \right\vert^{k} \right\rangle^{Bures}_{2-rebits}$--based upon 10,500,000 random realizations--to corresponding exact Hilbert-Schmidt moments. For $n=k=1$, the HS moment is zero. The HS moments themselves are symmetric under the interchange of $n$ and $k$, except when $n$ or $k$ equals zero.}
\end{figure}

\begin{figure}
\includegraphics{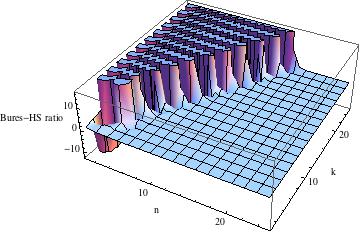}
\caption{\label{fig:RatioTwoQubitMoments2D}Ratio of numerically-generated {\it bivariate} Bures moments $\left\langle \left\vert \rho^{PT}\right\vert^{n} \left\vert \rho \right\vert^{k} \right\rangle^{Bures}_{2-qubits}$--based upon 10,500,000 random realizations--to corresponding exact Hilbert-Schmidt moments. The HS moments themselves are symmetric under the interchange of $n$ and $k$, except when $n$ or $k$ equals zero.}
\end{figure}

Further still, in
Figs.~\ref{fig:RatioTwoRebitMoments2D} and \ref{fig:RatioTwoQubitMoments2D} we plot the Bures-to-Hilbert-Schmidt ratios $\frac{\left\langle \left\vert \rho^{PT}\right\vert^{n}\left\vert 
\rho\right\vert^{k}\right\rangle^{Bures}}{\left\langle \left\vert \rho^{PT}\right\vert^{n}\left\vert 
\rho\right\vert^{k}\right\rangle^{HS}}, n, k = 0, 1, \ldots,24$ of the {\it bivariate} moments. Let us note, interestingly, that the Hilbert-Schmidt bivariate moments themselves are symmetric/invariant under the interchange of $n$ and $k$, except when $n$ or $k$ equals zero \cite[Apps. D.4, D.6]{MomentBased}. The corresponding 
formula \cite[App. D.6]{MomentBased} is
\begin{align*}
&  \frac{\left\langle \left\vert \rho^{PT}\right\vert ^{n}\left\vert \rho\right\vert
^{k}\right\rangle^{HS}_{\alpha}}{\left\langle \left\vert \rho\right\vert ^{k}\right\rangle^{HS}_{\alpha}}
  =\frac{\left(  k+1\right)  _{n}\left(  k+1+\alpha\right)  _{n}\left(
k+1+2\alpha\right)  _{n}}{2^{6n}\left(  k+3\alpha+\frac{3}{2}\right)
_{n}\left(  2k+6\alpha+\frac{5}{2}\right)  _{2n}}\\
&  \times~_{5}F_{4}\left(
\genfrac{}{}{0pt}{}{-n,-k,\alpha,\alpha+\frac{1}{2},-2k-2n-1-5\alpha
}{-k-n-\alpha,-k-n-2\alpha,-\frac{k+n}{2},-\frac{k+n-1}{2}}%
;1\right)  .
\end{align*}
where
\begin{equation} \label{generalformula}
\left\langle \left\vert \rho\right\vert ^{k}\right\rangle^{HS}_{\alpha} =\frac{k!\left(
\alpha+1\right)  _{k}\left(  2\alpha+1\right)  _{k}}{2^{6k}\left(
3\alpha+\frac{3}{2}\right)  _{k}\left(  6\alpha+\frac{5}{2}\right)  _{2k}}.
\end{equation}

\begin{figure}
\includegraphics{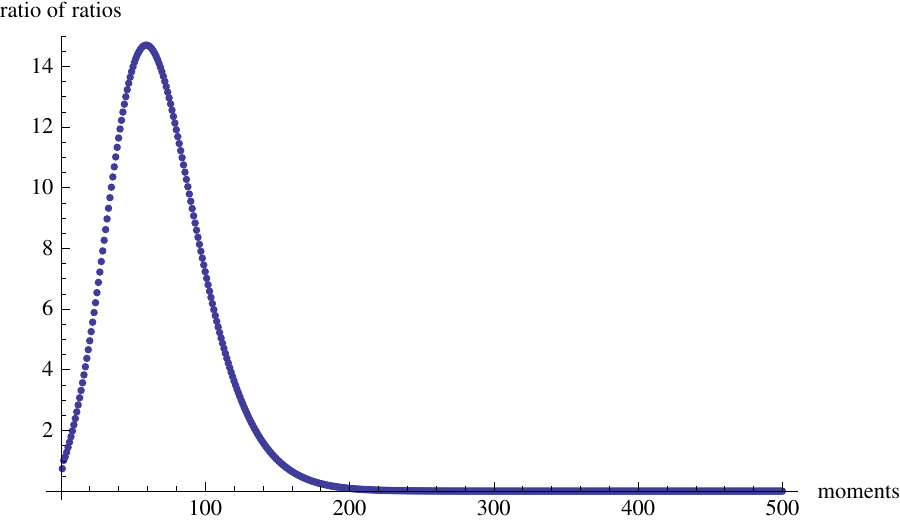}
\caption{\label{fig:RatioofRatios}Ratio of $\frac{\left\langle \left\vert \rho^{PT}\right\vert \left\vert \rho\right\vert
^{k}\right\rangle^{Bures}_{2-qubits}}{\left\langle \left\vert \rho\right\vert ^{k}\right\rangle^{Bures}_{2-rebits}}$ to $\frac{\left\langle \left\vert \rho^{PT}\right\vert \left\vert \rho\right\vert
^{k}\right\rangle^{HS}_{\alpha=1}}{\left\langle \left\vert \rho\right\vert ^{k}\right\rangle^{HS}_{\alpha=\frac{1}{2}}}$}
\end{figure}

In Fig.~\ref{fig:RatioofRatios} we display the (relatively well-behaved) ratio of the numerically-estimated Bures-ratio $\frac{\left\langle \left\vert \rho^{PT}\right\vert \left\vert \rho\right\vert
^{k}\right\rangle^{Bures}_{2-qubits}}{\left\langle \left\vert \rho\right\vert ^{k}\right\rangle^{Bures}_{2-rebits}}$ to the HS-ratio $\frac{\left\langle \left\vert \rho^{PT}\right\vert \left\vert \rho\right\vert
^{k}\right\rangle^{HS}_{\alpha=1}}{\left\langle \left\vert \rho\right\vert ^{k}\right\rangle^{HS}_{\alpha=\frac{1}{2}}}$.
\section{Bures Determinantal Moment Computatations} \label{Exact}
\subsection{Two-qubit results}
Now, let us present the following series 
($\left\langle  \left\vert  \rho^{PT}\right\vert  \left\vert  \rho \right\vert^{n}  \right\rangle_{2-qubits}^{Bures}$, $n =0, 1,\ldots 6$) of {\it exact} results (as inferred by us from high-precision computations),
\begin{equation} \label{firstmoment}
\left\langle \left\vert \rho^{PT}\right\vert \right\rangle_{2-qubits}^{Bures} = -\frac{1}{256} = - 2^{-8} \approx -0.00390625
\end{equation}
(the numerical estimate from the sample used to generate Fig.~\ref{fig:RatioTwoQubitMoments} being 
-0.003798959, while  the Hilbert-Schmidt counterpart is $\left\langle \left\vert \rho^{PT}\right\vert \right\rangle_{\alpha=1}^{HS} = -\frac{7}{3876} =-\frac{7}{2^2 \times 3 \times 17 \times 19}$),
\begin{equation}
\left\langle  \left\vert  \rho^{PT}\right\vert  \left\vert  \rho \right\vert  \right\rangle_{2-qubits}^{Bures} = -\frac{137}{1124597760} = - 
\frac{137}{2^{19} \times 3 \times 5 \times 11 \times 13},
\end{equation}
($\left\langle  \left\vert  \rho^{PT}\right\vert  \left\vert  \rho \right\vert  \right\rangle_{\alpha=1}^{HS} =  -\frac{1}{4576264}=-\frac{1}{2^3 \times 7 \times 11 \times 17 \times 19 \times 23}$)
\begin{equation}
\left\langle  \left\vert  \rho^{PT}\right\vert  \left\vert  \rho \right\vert^{2}  \right\rangle_{2-qubits}^{Bures} = -\frac{83}{4786288066560} = - 
\frac{83}{2^{24} \times 3 \times 5 \times 7 \times 11 \times 13 \times 19}
\end{equation}
\begin{equation}
\left\langle  \left\vert  \rho^{PT}\right\vert  \left\vert  \rho \right\vert^{3}  \right\rangle_{2-qubits}^{Bures} = - \frac{565}{219011789856178176} = -
\frac{565}{2^{36} \times 3 \times 11 \times 13 \times 17 \times 19 \times 23}
\end{equation}
\begin{equation}
\left\langle  \left\vert  \rho^{PT}\right\vert  \left\vert  \rho \right\vert^{4}  \right\rangle_{2-qubits}^{Bures} =  \frac{139}{52562829565482762240} 
= 
\frac{139}{2^{40} \times 3^2 \times 5 \times 11 \times 13 \times 17 \times 19 \times 23}
\end{equation}
and
\begin{equation} \label{lastmoment}
\left\langle  \left\vert  \rho^{PT}\right\vert  \left\vert  \rho \right\vert^{5}  \right\rangle_{2-qubits}^{Bures} =  \frac{14069}{1955073914750458962903040} 
= 
\frac{14069}{2^{52}  \times 5  \times 13 \times 17 \times 19 \times 23 \times 29 \times 31}.
\end{equation}
Having also found $\left\langle  \left\vert  \rho^{PT}\right\vert  \left\vert  \rho \right\vert^{k}  \right\rangle_{2-qubits}^{Bures}, k=6,\ldots,12 $, we were able to deduce 
(using the FindSequenceCommand of Mathematica) the underlying explanatory relation
\begin{equation} \label{explanatory}
\frac{\left\langle \left\vert \rho^{PT}\right\vert \left\vert \rho\right\vert
^{k}\right\rangle^{Bures}_{2-qubits}}{\left\langle \left\vert \rho\right\vert ^{k}\right\rangle^{Bures}_{2-qubits}}
=\frac{8 k^5+36 k^4-82 k^3-681 k^2-1366 k-885}{128 \left(16 k^5+192 k^4+883 k^3+1947
   k^2+2062 k+840\right)} = 
\end{equation}
\begin{displaymath}
\frac{k (k (2 k (2 k (2 k+9)-41)-681)-1366)-885}{128 (k+2) (k+3) (k+4) (4 k+5) (4 k+7)},
\end{displaymath}
where
\begin{equation} \label{BuresDet}
\left\langle   \left\vert  \rho \right\vert^{k}  \right\rangle_{2-qubits}^{Bures}
=\frac{315\ 2^{-8 k-1} \Gamma \left(k+\frac{1}{2}\right) \Gamma
   \left(k+\frac{3}{2}\right) \Gamma (2 k+2)}{\sqrt{\pi } \Gamma (k+3)
   \Gamma (k+4) \Gamma \left(2 k+\frac{9}{2}\right)}.
\end{equation}
The (lower-degree) Hilbert-Schmidt counterpart to 
the fundamental result (\ref{explanatory}) is \cite[eq. (24)]{MomentBased}
\begin{equation}
\frac{\left\langle \left\vert \rho^{PT}\right\vert \left\vert \rho\right\vert
^{k}\right\rangle^{HS}_{\alpha=1}}{\left\langle \left\vert \rho\right\vert ^{k}\right\rangle^{HS}_{\alpha=1}} = 
\frac{k^3+6 k^2-k-42}{256 k^3+3456 k^2+15536 k+23256} =
\frac{k (k (k+6)-1)-42}{8 (2 k+9) (4 k+17) (4 k+19)},
\end{equation}
where (substituting $\alpha =1$ into (\ref{generalformula}))
\begin{equation}
\left\langle \left\vert \rho\right\vert ^{k}\right\rangle^{HS}_{\alpha=1} = 
\frac{108972864000 \Gamma (k+1) \Gamma (k+2) \Gamma (k+3)
   \Gamma (k+4)}{\Gamma (4 (k+4))}. 
\end{equation}

In developing this last series of exact results (\ref{firstmoment})-(\ref{explanatory}), we combined symbolic methods with high-precision numerical computations as to the Bures expected values of monomials of the 
form \
\begin{equation} \label{monomialformula}
\lambda_{1}^{k_{1}} \lambda_{2}^{k_{2}} \lambda_{3}^{k_{3}} \lambda_{4}^{k_{4}},
\end{equation}
where the $\lambda$'s are the eigenvalues of $\rho$ (with, of course, $\lambda_{4}= 1- \lambda_{1}- \lambda_{2}- \lambda_{3}$). 
If the monomial is fully symmetric 
(that is $k_1=k_2=k_3=k_4$), then--since $|\rho| = 
\lambda_{1} \lambda_{2}  \lambda_{3} \lambda_{4}$--we can directly rely upon the {\it determinantal} moments implicit in the formulas \cite{szBures,csz}. Otherwise, unfortunately--providing the greatest impediment  to the research program pursued here--no general formulas seemed  at hand. We initially, thus, had to proceed with the somewhat tedious process of numerically analyzing, to high precision, each specific monomial 
arising in our computations. For instance, in computing the two-qubit result (\ref{explanatory}) 
we have used (from \cite{szBures}) $\left\langle  \lambda_{1}  \lambda_{2} \lambda_{3} \lambda_{4} \right\rangle_{2-qubits}^{Bures}= \frac{1}{16896}= (2^9 \times 3 \times 11)^{-1} $, as well as (our findings)
\begin{equation} \label{fourformulas}
\left\langle  \lambda_{1}^2  \lambda_{2} \lambda_{3}   \right\rangle_{2-qubits}^{Bures}=
\frac{43}{50688}, \left\langle  \lambda_{1}^2  \lambda_{2}^2 \right\rangle_{2-qubits}^{Bures}= 
\frac{83}{16896}, 
\end{equation}
\begin{displaymath}
\left\langle  \lambda_{1}^3  \lambda_{2} \right\rangle_{2-qubits}^{Bures}= 
\frac{457}{50688}, \left\langle  \lambda_{1}^4  \right\rangle_{2-qubits}^{Bures} =\frac{1127}{16896}.
\end{displaymath}
However, we did succeed (again employing the Mathematica FindSequenceFunction command) in generalizing (sufficient for our needs in computing the explanatory formula 
(\ref{explanatory})) this last set of four results to 
\begin{equation}
\frac{\langle \lambda_1^{k+1} \lambda_2^k \lambda_3^k \lambda_4^{k-1}\rangle^{Bures}_{2-qubits}}{\langle |\rho|^k \rangle^{Bures}_{2-qubits}} = 
\frac{\langle \lambda_1^{2} \lambda_2 \lambda_3  |\rho|^{k-1} \rangle^{Bures}_{2-qubits}}{\langle |\rho|^k \rangle^{Bures}_{2-qubits}}=
\frac{4 k (k+5)+19}{4 k^2-1},
\end{equation}
where $\langle |\rho|^k \rangle^{Bures}_{2-qubits} = \langle \lambda_{1}^{k} \lambda_{2}^{k} \lambda_{3}^{k} \lambda_{4}^{k} \rangle^{Bures}_{2-qubits}$ is given by (\ref{BuresDet}).
Further,
\begin{equation}
\frac{\langle \lambda_1^{k+1} \lambda_2^{k+1} \lambda_3^{k-1} \lambda_4^{k-1}\rangle^{Bures}_{2-qubits}}{\langle |\rho|^k \rangle^{Bures}_{2-qubits}} = 
\frac{k (k (4 k (k+14)+279)+556)+350}{k (k+4) (2 k-1) (2 k+1)},
\end{equation}
\begin{equation}
\frac{\langle \lambda_1^{k+2} \lambda_2^k \lambda_3^{k-1} \lambda_4^{k-1}\rangle^{Bures}_{2-qubits}}{\langle |\rho|^k \rangle^{Bures}_{2-qubits}} = \frac{k (k (4 k (k+19)+439)+996)+770}{k (k+4) (2 k-1) (2 k+1)},
\end{equation}
and
\begin{equation}
\frac{\langle \lambda_1^{k+3} \lambda_2^{k-1} \lambda_3^{k-1} \lambda_4^{k-1}\rangle^{Bures}_{2-qubits}}{\langle |\rho|^k \rangle^{Bures}_{2-qubits}} = \frac{(2 k+5) (k (k (4 k (k+32)+1067)+3148)+2898)}{k (k+4) (2 k-1) (2
   k+1)^2}.
\end{equation}
\subsection{Two-rebit results}
Turning from the generic (15-dimensional) two-bit case to the 
(9-dimensional) two-rebit one, we have found strong evidence that 
\begin{equation} \label{biggy}
\left\langle \left\vert \rho^{PT}\right\vert \right\rangle_{2-rebits}^{Bures} = -\frac{2663}{860160} = - \frac{2663}{2^{13} \times 3 \times 5 \times 7} \approx -0.00309594.
\end{equation}
The estimate for this quantity from the sample used to generate Fig.~\ref{fig:RatioTwoRebitMoments} is 
-0.00331101256, while  $\left\langle \left\vert \rho^{PT}\right\vert \right\rangle_{\alpha=\frac{1}{2}}^{HS} =-\frac{1}{858} =-\frac{1}{2 \times 3 \times 11 \times 13}$ \cite[App. A]{MomentBased}. (It might be noted that a quite close approximation to the result 
(\ref{biggy})--but one we reject--is the reciprocal integer 
$-\frac{1}{323}= -\frac{1}{17 \times 19} \approx -0.00309597$.)
In computing this two-rebit result
we have used (from \cite{szBures}) $\left\langle  \lambda_{1}  \lambda_{2} \lambda_{3} \lambda_{4} \right\rangle_{2-rebits}^{Bures}= \frac{1}{8192} =2^{-13}$, as well as our additional expected-value-monomial findings (cf. (\ref{fourformulas}))
\begin{equation}
\left\langle  \lambda_{1}^2  \lambda_{2} \lambda_{3}   \right\rangle_{2-rebits}^{Bures}=
\frac{41}{40960}, \left\langle  \lambda_{1}^2  \lambda_{2}^2 \right\rangle_{2-rebits}^{Bures}= 
\frac{1399}{286720}, \left\langle  \lambda_{1}^3  \lambda_{2} \right\rangle_{2-rebits}^{Bures}= 
\frac{2507}{286720}, \left\langle  \lambda_{1}^4  \right\rangle_{2-rebits}^{Bures} =\frac{18463}{286720}.
\end{equation}

The computation to obtain (\ref{biggy}) then took the form
\begin{equation}
\left\langle \left\vert \rho^{PT}\right\vert \right\rangle_{2-rebits}^{Bures} = -\frac{2663}{860160} =
\end{equation}
\begin{displaymath}
 \frac{26}{75}\langle \lambda_1^2 \lambda_2 \lambda_3 \rangle^{Bures}_{2-rebits} +\frac{53}{300}
   \langle \lambda_1^2 \lambda_2^2  \rangle^{Bures}_{2-rebits}-\frac{2}{15} \langle \lambda_1^3 \lambda_2  \rangle^{Bures}_{2-rebits}-\frac{1}{20}
  \langle \lambda_1^4  \rangle^{Bures}_{2-rebits}+\frac{33}{409600},
\end{displaymath}
where the constant term $\frac{33}{409600} =\frac{33}{50} \langle \lambda_1 \lambda_2 \lambda_3 \lambda_4 \rangle^{Bures}_{2-rebits}$.

Further, developing a two-rebit series analogous to the two-qubit series (\ref{firstmoment})-(\ref{lastmoment}), we have
\begin{equation}
\left\langle  \left\vert  \rho^{PT}\right\vert  \left\vert  \rho \right\vert  \right\rangle_{2-rebits}^{Bures} = -\frac{127}{1132462080}=-
\frac{127}{2^{23} \times 3^3 \times 5}
\end{equation}
(while $\left\langle  \left\vert  \rho^{PT}\right\vert  \left\vert  \rho \right\vert  \right\rangle_{\alpha=\frac{1}{2}}^{HS} = 0$),
\begin{equation}
\left\langle  \left\vert  \rho^{PT}\right\vert  \left\vert  \rho \right\vert^{2}  \right\rangle_{2-rebits}^{Bures} = \frac{1257}{105827994173440} = - 
\frac{3 \times 419}{2^{38}  \times 5 \times 7 \times 11}
\end{equation}
\begin{equation}
\left\langle  \left\vert  \rho^{PT}\right\vert  \left\vert  \rho \right\vert^{3}  \right\rangle_{2-rebits}^{Bures} = \frac{13813}{221661544159641600} = - 
\frac{19 \times 727}{2^{47} \times 3^2 \times 5^2 \times 7},
\end{equation}
and 
\begin{equation}
\left\langle  \left\vert  \rho^{PT}\right\vert  \left\vert  \rho \right\vert^{4}  \right\rangle_{2-rebits}^{Bures} = \frac{3927}{28823037615171174400} = 
\frac{3 \times 7 \times 11 \times 17}{2^{60} \times 5^2}.
\end{equation}
Then, having found the next nine  members of this series, we were able to generate the explanatory formula for the infinite series
(cf. the two-qubit result (\ref{explanatory}), also the ratio of degree-5 polynomials in $k$)
\begin{equation} \label{RebitExplanatory}
\frac{\left\langle \left\vert \rho^{PT}\right\vert \left\vert \rho\right\vert
^{k}\right\rangle^{Bures}_{2-rebits}}{\left\langle \left\vert \rho\right\vert ^{k}\right\rangle^{Bures}_{2-rebits}}
= \frac{64 k^5+128 k^4-340 k^3-1032 k^2-1099 k-384}{k \left(8 k^2-2
   k-1\right) \left(8 k^2+18 k-5\right)} =
\end{equation}
\begin{displaymath}
\frac{k (4 k (k (16 k (k+2)-85)-258)-1099)-384}{k (2 k-1) (2 k+5) (4 k-1)
   (4 k+1)},
\end{displaymath}
where
\begin{equation} \label{BuresDetRebit}
\left\langle   \left\vert  \rho \right\vert^{k}  \right\rangle_{2-rebits}^{Bures}
= \frac{3 2^{2-8 k} \Gamma \left(2 k+\frac{3}{2}\right)}{\sqrt{\pi }
   \left(2 k^2+3 k+1\right) \Gamma (2 k+4)}.
\end{equation}
The (simpler, lower-degree) Hilbert-Schmidt counterpart to 
(\ref{RebitExplanatory}) is
\cite[eq. 3]{MomentBased}
\begin{equation}
\frac{\left\langle \left\vert \rho^{PT}\right\vert \left\vert \rho\right\vert
^{k}\right\rangle^{HS}_{\alpha=\frac{1}{2}}}{\left\langle \left\vert \rho\right\vert ^{k}\right\rangle^{HS}_{\alpha=\frac{1}{2}}} = 
\frac{(k-1) (k (2 k+11)+16)}{32 (k+3) (4 k+11) (4 k+13)}.
\end{equation}
It does appear that in the Bures case, such formulas are the ratios of
degree-$5 n$ polynomials in $k$, while in the Hilbert-Schmidt case they are ratios 
of degree-$3 n$ polynomials in $k$ \cite{MomentBased}. This "higher-degreeness" might be taken as some indication of the greater complexity of the Bures $2 \times 2$ determinantal moment problem (cf. (\ref{HSformula1}),  
(\ref{Buresformula1})).

As intermediate steps in deriving 
(\ref{RebitExplanatory}), we have developed the formulas 
\begin{equation} \label{earlier}
\frac{\langle \lambda_1^{k+1} \lambda_2^k \lambda_3^k \lambda_4^{k-1}\rangle^{Bures}_{2-rebits}}{\langle |\rho|^k \rangle^{Bures}_{2-rebits}} =  
\frac{\langle \lambda_1^{2} \lambda_2 \lambda_3  |\rho|^{k-1} \rangle^{Bures}_{2-rebits}}{\langle |\rho|^k \rangle^{Bures}_{2-rebits}}
\end{equation}
\begin{displaymath}
=\frac{8 k^2+22 k+11}{8 k^2-2 k-1} = -\frac{4}{4 k+1}+\frac{8}{2 k-1}+1,
\end{displaymath}
\begin{equation}
\frac{\langle \lambda_1^{k+1} \lambda_2^{k+1} \lambda_3^{k -1}\lambda_4^{k-1}\rangle^{Bures}_{2-rebits}}{\langle |\rho|^k \rangle^{Bures}_{2-rebits}} = \frac{4 k (k (16 k (k+8)+363)+406)+545}{(2 k-1) (2 k+5) (4 k-1) (4 k+1)},
\end{equation}
\begin{equation}
\frac{\langle \lambda_1^{k+2} \lambda_2^k \lambda_3^{k-1} \lambda_4^{k-1}\rangle^{Bures}_{2-rebits}}{\langle |\rho|^k \rangle^{Bures}_{2-rebits}} = \frac{4 k (k (16 k (k+11)+591)+766)+1325}{(2 k-1) (2 k+5) (4 k-1) (4
   k+1)},
\end{equation}
and 
\begin{equation} \label{later}
\frac{\langle \lambda_1^{k+3} \lambda_2^{k-1} \lambda_3^{k-1} \lambda_4^{k-1}\rangle^{Bures}_{2-rebits}}{\langle |\rho|^k \rangle^{Bures}_{2-rebits}} =\frac{k (4 k (k (16 k (k+20)+1899)+4678)+20057)+7680}{k (2 k-1) (2 k+5)
   (4 k-1) (4 k+1)}.
\end{equation}

Further,
\begin{equation} \label{biggy2}
\left\langle \left\vert \rho^{PT}\right\vert^2 \right\rangle_{2-rebits}^{Bures} =  \frac{50654227}{1307993702400} =
\frac{13 \times 101 \times 173 \times 223}{2^{23} \times 3^4 \times 5^2 \times 7 \times 11} \approx 0.0000387267.
\end{equation}
(The sample [Ginibre-ensemble] estimate--from Fig.~\ref{fig:RatioTwoRebitMoments}--was 0.000036949.)
In attempting to find the $n=2$ analogue of (\ref{RebitExplanatory}), we have computed
\begin{equation} \label{uuu}
\frac{\langle \lambda_1^3 \lambda_2^2 \lambda_3^2 \lambda_4 |\rho|^k \rangle_{2-rebits}^{Bures}}{\langle |\rho|^{k+2} \rangle_{2-rebits}^{Bures}} = \frac{8 k^2 +54 k +87}{8 k^2 +30 k +27}.
\end{equation}
(The right-hand side of this equation can be obtained  by replacing $k$ in (\ref{earlier}) by $k+2$.)
Also,
\begin{equation} \label{uuu2}
\frac{\langle \lambda_1^3 \lambda_2^3 \lambda_3^2  |\rho|^k \rangle_{2-rebits}^{Bures}}{\langle |\rho|^{k+2} \rangle_{2-rebits}^{Bures}} = \frac{128 k^5+2496 k^4+19000 k^3+70284 k^2+125922 k+87213}{128 k^5+1344 k^4+5080
   k^3+8700 k^2+6642 k+1701}= 
\end{equation}
\begin{displaymath}
\frac{176}{63 (4 k+9)}-\frac{334}{6 k+9}+\frac{495}{14 k+7}+\frac{1260}{44
   k+77}-\frac{5}{99 (2 k+9)}+1.
\end{displaymath}
We required fifteen distinct monomial-expected-value computations on the order of (\ref{uuu}) and (\ref{uuu2}). Four of them could be directly obtained from 
(\ref{earlier})-(\ref{later}) by the replacement of $k$ by $k+2$, and one immediately from the expected determinant formula, while the other ten required apparently new computations. We have been able to determine six of these ten so far. Among the successful computations were:
\begin{equation}
\frac{\langle \lambda_1^4 \lambda_2^3 \lambda_3  |\rho|^k \rangle_{2-rebits}^{Bures}}{\langle |\rho|^{k+2} \rangle_{2-rebits}^{Bures}} = \frac{512 k^6+13696 k^5+143008 k^4+751304 k^3+2096676 k^2+2936118 k+1596285}{(4 k+5)
   \left(128 k^5+1344 k^4+5080 k^3+8700 k^2+6642 k+1701\right)}= 
\end{equation}
\begin{displaymath}
\frac{5}{858 (2 k+9)}+\frac{675}{4 k+2}-\frac{184}{3 (4 k+9)}+\frac{6480}{44
   k+77}-\frac{14700}{52 k+65}-\frac{6}{2 k+3}+1,
\end{displaymath}
and
\begin{equation}
\frac{\langle \lambda_1^5 \lambda_2^2 \lambda_3  |\rho|^k \rangle_{2-rebits}^{Bures}}{\langle |\rho|^{k+2} \rangle_{2-rebits}^{Bures}} = 
\end{equation}
\begin{displaymath}
\frac{512 k^7+17792 k^6+234144 k^5+1576840 k^4+5978260 k^3+12867318 k^2+14589933
   k+6699810}{(k+2) (4 k+5) \left(128 k^5+1344 k^4+5080 k^3+8700 k^2+6642
   k+1701\right)} =
\end{displaymath}
\begin{displaymath}
-\frac{175}{2574 (2 k+9)}-\frac{122500}{39 (4 k+5)}+\frac{17492}{63 (4
   k+9)}-\frac{1936}{6 k+9}+\frac{55740}{44 k+77}+\frac{32765}{84 k+42}-\frac{1792}{3
   (k+2)}+1.
\end{displaymath}
The Hilbert-Schmidt two-qubit counterpart of this last formula is
\begin{equation}
\frac{\langle \lambda_1^5 \lambda_2^2 \lambda_3  |\rho|^k \rangle_{2-qubits}^{HS}}{\langle |\rho|^{k+2} \rangle_{2-qubits}^{HS}} = 
\frac{\langle \lambda_1^5 \lambda_2^2 \lambda_3  |\rho|^k \rangle_{\alpha=1}^{HS}}{\langle |\rho|^{k+2} \rangle_{\alpha=1}^{HS}}
\end{equation}
\begin{displaymath}
   \frac{k^6+50 k^5+851 k^4+6770 k^3+27234 k^2+52970 k+39084}{64 \left(1024
   k^6+30720 k^5+383104 k^4+2542080 k^3+9465796 k^2+18753960
   k+15444891\right)}
\end{displaymath}
\begin{displaymath}
= \frac{(k+2) (k+3) (k (k (k (k+45)+620)+3400)+6514)}{64 (2 k+9) (2 k+11)
   (4 k+17) (4 k+19) (4 k+21) (4 k+23)}.
\end{displaymath}
The four formulas remaining to be determined are those for 
$\frac{\langle \lambda_1^5 \lambda_2^3  |\rho|^k \rangle_{2-rebits}^{Bures}}{\langle |\rho|^{k+2} \rangle_{2-rebits}^{Bures}}, \frac{\langle \lambda_1^6 \lambda_2^2  |\rho|^k \rangle_{2-rebits}^{Bures}}{\langle |\rho|^{k+2} \rangle_{2-rebits}^{Bures}},
\frac{\langle \lambda_1^7 \lambda_2   |\rho|^k \rangle_{2-rebits}^{Bures}}{\langle |\rho|^{k+2} \rangle_{2-rebits}^{Bures}}$ and 
$\frac{\langle \lambda_1^8  |\rho|^k \rangle_{2-rebits}^{Bures}}{\langle |\rho|^{k+2} \rangle_{2-rebits}^{Bures}}$.
\subsection{Computations for generic $3 \times 3$ real density matrices}
As a supplementary exercise in finding the expected values of monomials of eigenvalues for generic real $n \times n$ density matrices ($\rho$), 
we have been able to determine that for $n=3$,
\begin{equation}
\frac{{\langle \lambda_1^{k+1} \lambda_2^k \lambda_3^{k-1} \rangle}_{retrit}^{Bures}}{\left\langle   \left\vert  \rho \right\vert^{k}  \right\rangle_{retrit}^{Bures}} = \frac{{\langle \lambda_1^{2} \lambda_2 |\rho|^{k-1} \rangle}_{retrit}^{Bures}}{\left\langle   \left\vert  \rho \right\vert^{k}  \right\rangle_{retrit}^{Bures}} =
\frac{16 k^2+36 k+13}{2 \left(8 k^2-2 k-1\right)} = -\frac{5}{12 k+3}+\frac{35}{6 (2 k-1)}+1, 
\end{equation}
and
\begin{equation}
\frac{{\langle \lambda_1^{k+2} \lambda_2^{k-1} \lambda_3^{k-1} \rangle}_{retrit}^{Bures}}{\left\langle   \left\vert  \rho \right\vert^{k}  \right\rangle_{retrit}^{Bures}} = \frac{32 k^3+224 k^2+308 k+101}{32 k^3-16 k^2-2 k+1}= \frac{4 k (8 k (k+7)+77)+101}{(2 k-1) (4 k-1) (4 k+1)}, 
\end{equation}
where
\begin{equation}
\left\langle   \left\vert  \rho \right\vert^{k}  \right\rangle_{retrit}^{Bures}= 
{\langle \lambda_1^{k} \lambda_2^k \lambda_3^{k} \rangle}_{retrit}^{Bures} = \frac{2^{1-8 k} \Gamma (4 k+2)}{\Gamma (3 (k+1)) \Gamma (k+2)}.
\end{equation}
\section{Discussion} \label{Discussion}
In our various Bures computations above, we have employed Euler-angle parameterizations of the two-rebit \cite[App. I]{JMP2008} and two-qubit density matrices 
\cite{tbs} (while in the Hilbert-Schmidt study \cite{MomentBased},  Cholesky-decomposition parameterizations were 
utilized).

We note that in the three-dimensional simplex of eigenvalues ($\lambda_{i} \geq 0, \Sigma_{i=1}^4 \lambda_i =1$), the Hilbert-Schmidt measure is proportional ($\alpha =\frac{1}{2}$ denoting the two-rebit systems, and $\alpha=1$, 
the two-qubit systems) to \cite[eq. (4.1)]{szHS}  
\begin{equation} \label{HSformula1}
P^{HS}_{\alpha} \propto \Pi_{i<j}^{4} |\lambda_i -\lambda_j|^{2 \alpha},
\end{equation}
while \cite[eq. (3.19)]{szBures}
\begin{equation} \label{Buresformula1}
P^{Bures}(\alpha) \propto \frac{\Pi_{i<j}^{4} \Big(\frac{{|\lambda_i -\lambda_j|}^2}{\lambda_i+\lambda_j}\Big)^{\alpha}}{\sqrt{\Pi_{i=1}^{4} \lambda_i}}.
\end{equation}
The normalization constants are, respectively, $80640$ and $378378000$
in the two HS cases of interest here, and $\frac{128}{\pi}$ and $\frac{71680}{\pi^2}$
for the two Bures cases, where the first member of each pair corresponds to 
$\alpha= \frac{1}{2}$ and the second member to $\alpha=1$. 

Further, it appears that the Hilbert-Schmidt determinantal moment formulas developed in \cite{MomentBased} potentially provide certain
"leverage" in obtaining (the desired) comparable Bures determinantal moment formulas. That is, if  one had
formulas for the expected values of the eigenvalue-monomials 
(\ref{monomialformula})
in {\it both} the Bures and Hilbert-Schmidt cases,
then one could reduce the computations for the Bures determinantal moments (avoiding or minimizing integrations  over the [more numerous] non-eigenvalue parameters) by solving 
certain linear systems that incorporate the Hilbert-Schmidt counterparts. 

A referee has indicated that it would be of considerable interest to conduct similar analyses to those here for $3 \times 3$-dimensional systems, in particular in light of certain recent work \cite{bert,baum} pertaining to them.

\begin{acknowledgments}
I would like to express appreciation to the Kavli Institute for Theoretical
Physics (KITP)
for computational support in this research. Michael Trott provided guidance regarding numerical integration procedures in Mathematica.
\end{acknowledgments}

\bibliography{BuresAttempt3}

\end{document}